\documentclass[reprint,amsmath,amssymb,aps,]{revtex4-1}
\usepackage{graphicx}
\usepackage{dcolumn}
\usepackage{bm}
\newcommand{\Tr}{\text{Tr}}
\begin{document}
\title{ Quantum speed-up based on classical-field and moving-qubit}
\author{Jianhe Yang$^\ddag$\textsuperscript{1}}
\thanks{$^\ddag$These authors contribute equally to this article.}
\author{Rongfang Liu$^\ddag$\textsuperscript{1}}%
\author{Hong-Mei Zou\textsuperscript{1}}%
\email{zhmzc1997@hunnu.edu.cn}
\author{Danping Lin\textsuperscript{2}}
\author{Ali Mortezapour\textsuperscript{3}}
\affiliation{$^1$Synergetic Innovation Center for Quantum Effects and Application, Key Laboratory of Low-dimensional Quantum Structures and Quantum Control of Ministry of Education, School of Physics and Electronics, Hunan Normal University, Changsha,410081, P.R. China\\
$^2$ Faculty of Science, Guilin University of Aerospace Technology, Guilin 541004,
People’s Republic of China\\
$^3$Department of Physics, University of Guilan, PO Box 41335–1914, Rasht, Iran}%

\begin{abstract}
In this work, we provide a model of a moving-qubit interacting with the multimode cavity, where the qubit is driven by the classical field. We obtain the analytic solution of the density operator of the qubit,  then investigate the quantum speed limit time(QSLT) and the non-Markovianity based on the classical field and the moving-velocity. The results show that the transition from Markovian to non-Markovian dynamics is the intrinsic physical reason of the quantum speed-up process, both of the driving field and the strong coupling can enhance the non-Markovianity in the dynamics process and speed up the evolution of the qubit, but the moving velocity of the qubit can decrease the non-Markovianity in dynamics process and delay the evolution of qubit. To some extent, the classical field can reduce the effect of the moving velocity of the qubit on the quantum evolution process.
\begin{description}
\item[PACS numbers]
03.65.Yz, 03.67.-a, 03.67.Lx
\item[Keywords]
quantum speed-up process, classical-field, moving-qubit
\end{description}

\end{abstract}

\maketitle

\section{Introduction}

Quantum mechanics restriction on the evolution speed of quantum systems is called quantum speed limit, this is the fundamental laws of nature\cite{S.Lloyd13,L.Mandelstam3,J.Ananda3,L.Vaidman3,S.Luo3,Breuer3,Michael3}. It has recently attracted considerable attention and played remarkable roles in various areas of quantum physics including quantum communication\cite{M.H.Yung3,M.A.Nielsen3,J.D.Bekenstein3}, quantum optimal control\cite{T.Caneva3,V.Mukherjee3,G.C.Hegerfeldt3,G.C.Hegerfeldt13,C.Avinadav3}, quantum computation\cite{S.Lloyd3,L.B.Levitin3} and nonequilibrium thermodynamics\cite{S.Deffner3}. In recent years, the quantum speed limit time and the non-Markovian dynamic process of an open quantum system have been widely concerned. Quantum speed limit(QSL) can effectually characterize the maximal speed of evolution of a quantum system from a given initial state to a target state\cite{J.Anandan}. Quantum speed limit time (QSLT) is defined as the minimal evolution time of a quantum system. For a unitary process, there are two common bounds of the QSLT. One is expressed as $\tau_{qsl}=\pi\hbar/(2\triangle E)$, where $\triangle E$ represents the energy fluctuation of the initial state, which is proposed by Mandels and Tamm (the MT bound).  The other is $\tau_{qsl}=\pi\hbar/(2\langle E\rangle)$, where $\tau_{qsl}$ depends on the average energy $\langle E\rangle$, which is derived by Margolus and Levitin (the ML bound). By combining the two bounds, the QSLT of the two orthogonal pure states in the closed system is $\tau_{qsl}=\max\{\pi\hbar/2\triangle E,\pi\hbar/2\langle E\rangle\}$\cite{Mandelstam, G.N.Fleming, J.D.Bekenstein, K.Bhattacharyya, L.Vaidman, Margolus}. The MT-QSL bound based on the relative purity, the ML-QSL and NI-QSL dependent on initial states as well as the quantum speed limit in a nonequilibrium environment have also been investigated in succession\cite{Campo,S.-X.Wu1,X.Cai,M.Schiro,F.Peronaci,P.Bhupathi,S.Oviedo-Casado,F.C.Lombardo}. N. Mirkin $et\ al.$ investigate the QSL bound in terms of the quantum Fisher information, different operator norms and the notion of quantumness, respectively\cite{Mirkin}. For an open system, one often uses the non-Markovianity to quantify the non-Markovian effect of its dynamical behavior. The measure of the non-Markovianity in quantum processes for an open two-level system has been presented in Refs.\cite{Laine,Zeng,He}, including the non-Markovianity in the dynamics process of an open quantum system\cite{Fanchini}, the non-Markovian character of colored noisy channels\cite{Benedtti}.  In particular, Deffner and Lutz acquires the unified bound of an open system by using the Bures angle based on the ML and MT bounds, and their result shows that the non-Markovian effects could speed up the quantum evolution\cite{Deffner1}.  In addition, many valuable effort have also been devoted to the relationships between the non-Markovianity and the QSL, such as quantum speedup in a memory environment\cite{Z.Y.Xu} and quantum speedup in open quantum systems\cite{H.B.Liu}, and so on. Y. J. Zhang and W. Han propose a method of accelerating the speed of evolution of an open system by an external classical driving field of a qubit in a zero-temperature structured reservoir\cite{Y.J.Zhang}.  
 
Most of the researches are based on quantum models of stationary qubits, but in the present experiments such as cavity QED, cooling of an atom does not make the atom completely stationary\cite{Zhang1,Zhang2,Zhang3}. Therefore, how does the velocity of moving-qubit effect the QSLT and the non-Markovianity? Can the external classical field regulate the QSLT and the non-Markovianity? These problems are also important and meaningful in experimental research of open quantum systems.

Authors in Ref.\cite{R} proposed a method of treatment of the spectrum of squeezing based on the modes of the universe, in which the mirrors at $z=l$ and $-L$ are completely reflective while the one at $z=0$ is semitransparent(Fig.1). A. Mortezapour and D. Park $et\ al$. also studied this model by assuming $l\rightarrow\infty$ and introducing the parameter $f(z)$ which describes the dependence function of qubit motion along the $z$-axis\cite{A.Mortezapour1,C.Leonardi,A.Mortezapour}. Inspired by these works and considered the actual experimental conditions, for example, $l$ is about $20cm$ path in  Ref.\cite{P}, we choose a parameter $l=23cm$ according to the experimental device. Meanwhile, we add an external classical field along the $y$-axis in order to effectively control the quantum effect of a moving-qubit. Though the physical model of a moving-qubit under the classical field in open system is very complicated, we choose dressed-state to simplify the solution process and obtain an analytical solution of this qubit by using some approximate conditions.

\begin{center}
	\includegraphics[width=8cm,height=3.5cm]{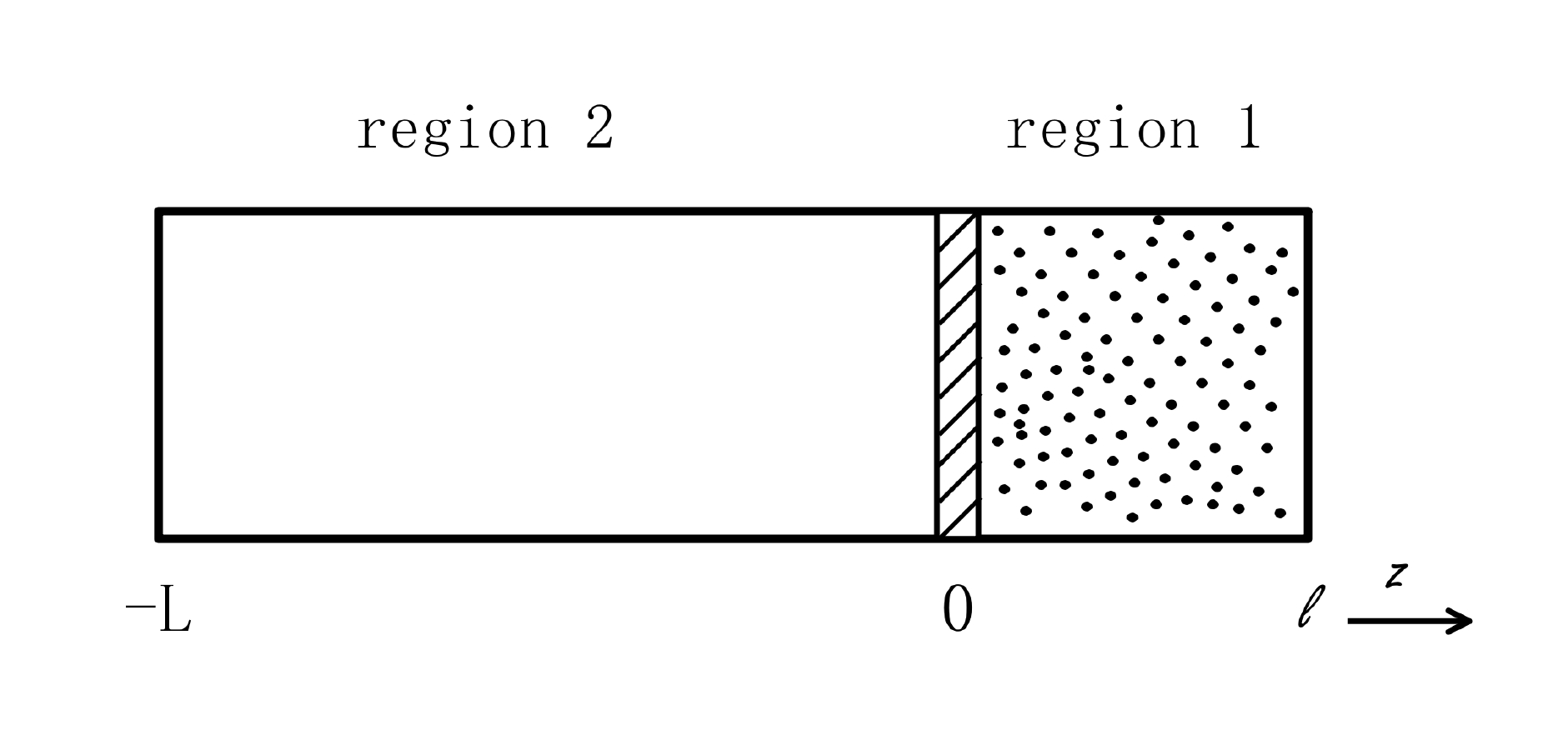}
	\parbox{8cm}{\small{\bf FIG.1.}
		Cavity with partially reflecting mirror imbedded in large ideal cavity($L\rightarrow\infty$).}
\end{center}

In this paper we investigated the quantum speed-up and the non-Markovianity based on the classical field and the velocity of the moving-qubit. We find that the classical field can accelerate the evolution of the qubit and increase the non-Markovianity of the dynamics process, but the velocity of the moving-qubit can delay the evolution of the qubit and decrease the non-Markovianity. Our results show that the QSLT of a moving-qubit can be effectively controlled by the driving strength and the qubit-velocity. This paper is organized as following. In Section \uppercase\expandafter{\romannumeral2}, we present a physical model and its analytical solution of a moving-qubit under the classical field. In section  \uppercase\expandafter{\romannumeral3}, we introduce the quantum speed limit and the non-Markovianity. In section  \uppercase\expandafter{\romannumeral4}, we give the results and discussions. Finally, we have a brief summary of our work in section  \uppercase\expandafter{\romannumeral5}.

\section{Physical model and analytical solution}

We consider a moving-qubit interacting with the multimode reservoir, where the qubit is driven by the classical field. Select parameter $l=23cm$ consistent with the experimental device and assuming $L\rightarrow\infty$. In this paper, the classical field is polarized along the $x$-axis and the qubit moves direction is along with the $z$-axis. Meanwhile, let the classical field along the $y$-axis in order to effectively control the quantum effect of the moving-qubit. The Hamiltonian reads ($\hbar=1$)

\begin{eqnarray}\label{EB01}
\begin{array}{l}
\begin{aligned}
\hat{H}=&\frac{1}{2}\omega_{0}\sigma_{z}+\sum_{k}\omega_{k}a_{k}^{\dag}a_{k}+\Omega e^{-i\omega_{L}t}\sigma_{+}+\Omega e^{i\omega_{L}t}\sigma_{-}\\
&+\sum_{k}g_{k}(f_{k}(z)a_{k}\sigma_{+}+f_{k}(z)a_{k}^{\dag}\sigma_{-}),
\end{aligned}
\end{array}
\end{eqnarray}
where the operators $\sigma_{z}$ and $\sigma_{\pm}$ are defined by $\sigma_{z}=|e\rangle\langle e|-|g\rangle\langle g|, \sigma_{+}=|e\rangle\langle g|$, and $\sigma_{-}=\sigma_{+}^{\dag}$ associated with the upper level $|e\rangle$ and lower level $|g\rangle$.  $\omega_{0}$ is the transition frequency of the qubit. $a_{k}^{\dag}(a_{k})$ and $\omega_{k}$ are the creation(annihilation) operator and the frequency of the $k$-th mode. In addition, $g_{k}$ denotes the coupling constant between the qubit and the $k$-th mode, $\Omega$ is the classical driving strength.  The parameter $f_{k}(z)$ describes the dependency function of the qubit motion along with the $z$-axis, and it is given by
\begin{eqnarray}\label{EB02}
\begin{aligned}
f_{k}(z)=f_{k}(vt)=\sin[k(z-l)]=\sin[\omega_{k}(\beta t-\tau_{0})],
\end{aligned}
\end{eqnarray}
where $\beta=\upsilon/c$ and $\tau_{0}=l/c$, $\upsilon$ and $c$ are respectively the velocities of the moving-qubit and the light, $l$ is the length of the right side cavity\cite{A.Mortezapour}. Note that the dependency function is not zero when $z=0$, while it is zero when $z=l$(perfect boundary).

\begin{center}
	\includegraphics[width=8cm,height=4.5cm]{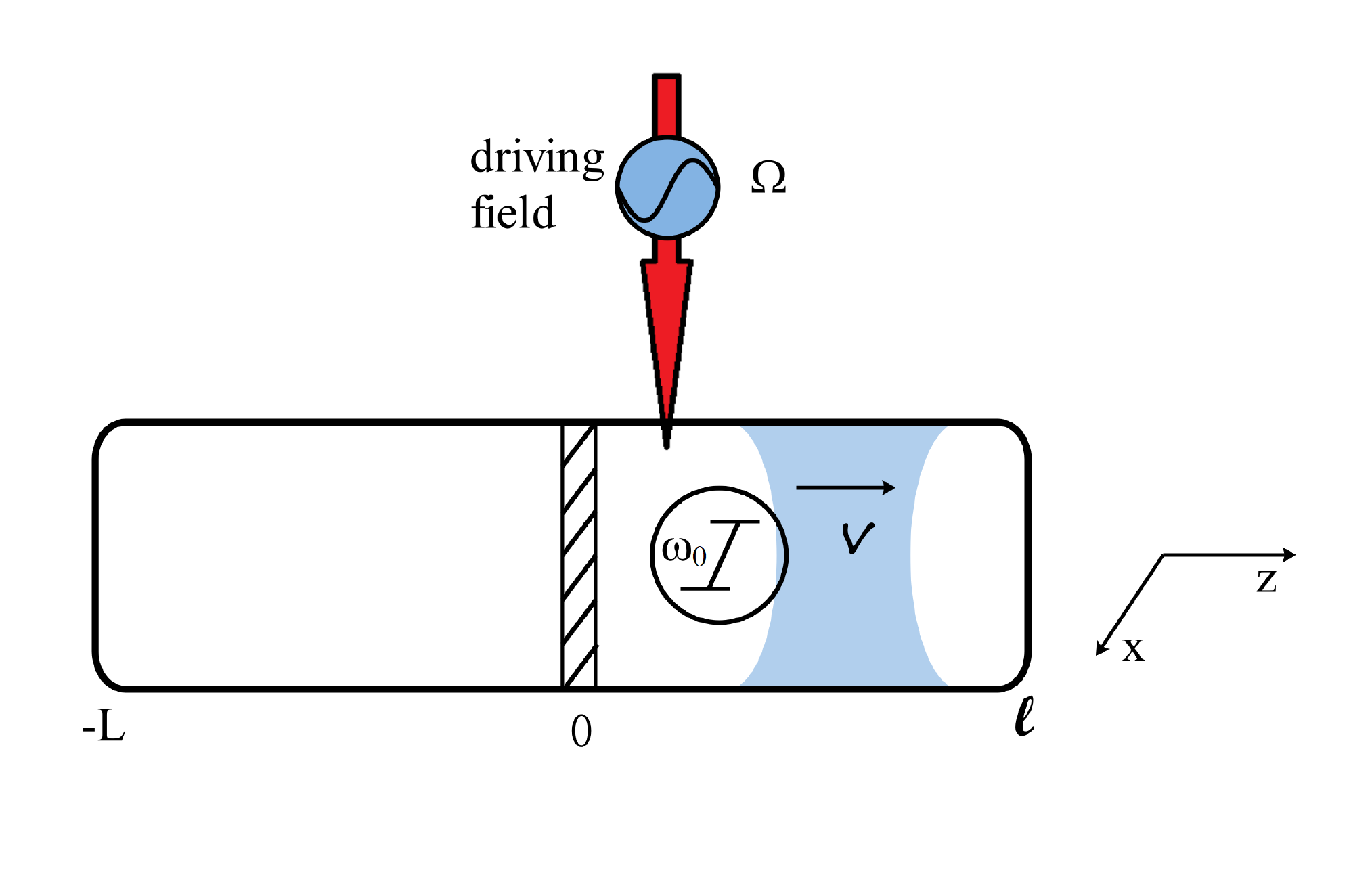}
	\parbox{8cm}{\small{\bf FIG.2.}
		Schematic illustration of a setup where a single qubit is moving inside a cavity and driven by the classical field. The qubit is a
		two-level atom with transition frequency $\omega_{0}$ traveling with constant velocity $v$.}
\end{center}

In the dressed-state basis $\{|E\rangle=\frac{1}{\sqrt{2}}(|g\rangle+|e\rangle), |G\rangle=\frac{1}{\sqrt{2}}(|g\rangle-|e\rangle)\}$,  by considering rotating reference frames through unitary transformation $U_{1}=\exp{[-i\omega_{L}\sigma_{z}t/2]}$ and $U_{2}=\exp{[-i\omega_{0}\Sigma_{z}t/2]}$,  four no-conservation energy terms would occur. By using the rotating-wave approximation, the Hamiltonian in  Eq. (\ref{EB01}) is equivalently transferred to an effective Hamiltonian
\begin{eqnarray}\label{EB03}
\begin{array}{l}
	\begin{aligned}
		\hat{H}_{e}=&\frac{\omega_{D}}{2}\Sigma_{z}+\sum_{k}\omega_{k}a_{k}^{\dag}a_{k}\\
		&+\sum_{k}g_{k}^{1}(f_{k}(z)a_{k}\Sigma_{+}+f_{k}(z)a_{k}^{\dag}\Sigma_{-}),
	\end{aligned}
\end{array}
\end{eqnarray}
where $\omega_{D}=\sqrt{\Delta^{2}+4|\Omega|^{2}}$, $g_{k}^{1}=g_{k}/2$, $\Delta=\omega_{0}-\omega_{L}$. $\Sigma_{z}=|E\rangle\langle E|-|G\rangle\langle G|, \Sigma_{+}=|E\rangle\langle E|$ and $\Sigma_{-}=\Sigma_{+}^{\dag}$.

In the interaction picture, the Hamiltonian reads
\begin{eqnarray}\label{EB04}
\begin{aligned}
\hat{H}_{I}=&\sum_{k}g_{k}^{1}[f_{k}(z)a_{k}\Sigma_{+}e^{i(\omega_{D}+\omega_{L}-\omega_{k})t}\\
&+f_{k}^{\ast}(z)a_{k}^{\dag}\Sigma_{-}e^{-i(\omega_{D}+\omega_{L}-\omega_{k})t}].
\end{aligned}
\end{eqnarray}

Furthermore, at zero temperature, consider that the transition frequency of the moving-qubit is resonant with the classical field frequency $(\omega_{0}-\omega_{L}=0)$ and suppose the initial state of the total system is  $|\psi(0)\rangle=C_{1}(0)|E\rangle|0\rangle$ , where $|0\rangle$ is denoting the vacuum state of the reservoir. At any time $t>0$ the state of the total system is given by  $|\psi(t)\rangle=C_{1}(t)|E\rangle|0\rangle+C_{0}(t)|G\rangle|0\rangle+\sum_{k}C_{k}(t)|G\rangle|1_{k}\rangle$, where the $|1_{k}\rangle$ is the state of the reservoir with only one excitation in the $k$-th mode. 
By solving the Schr\"{o}dinger equation , we can obtain

\begin{eqnarray}\label{EB05}
\begin{aligned}
&\dot{C}_{1} (t)=-i\sum_{k}g_{k}^{1}f_{k}(z)C_{k}(t)e^{i(\omega_{D}+\omega_{L}-\omega_{k}) t}\\
&\dot{C}_{k} (t)=-i\sum_{k} g_{k}^{1}f_{k}(z) C_{1}(t)e^{-i(\omega_{D}+\omega_{L}-\omega_{k})t}.
\end{aligned}
\end{eqnarray}
Owing to no excitations in the initial state of the reservoir, that is to say$C_{k}(0)=0$ , one can obtain from Eqs. (\ref{EB05})
\begin{eqnarray}\label{EB08}
\begin{array}{l}
\begin{aligned}
\dot{C}_{1}(t)=-\int_{0}^{t}dt_{1}F(t-t_{1})C_{1}(t_{1}),
\end{aligned}
\end{array}
\end{eqnarray}
the correlation function $F(t-t_{1})$ can be expressed as the following form
\begin{eqnarray}\label{EB09}
\begin{array}{l}
\begin{aligned}
F(t-t_{1})=&\int_{0}^{\infty}J(\omega_{k}) \sin[\omega_{k} (\beta t-\tau_{0})]\times\\
&\sin[\omega_{k}(\beta t_{1}-\tau_{0})] e^{i(\omega_{D}+\omega_{L}-\omega_{k})(t-t_{1})}d\omega_{k},
\end{aligned}
\end{array}
\end{eqnarray}
where $J(\omega)$ is the spectral density of the reservoir. If the structure of the reservoir has the Lorentzian form
\begin{eqnarray}\label{EB10}
J(\omega_{k})=\frac{1}{2\pi}\frac{\gamma \lambda^{2}}{[(\omega_{0}-\omega_{k})^{2}+\lambda^{2}]},
\end{eqnarray}
where $\lambda$ is the spectral width of the reservoir, $\gamma$ is the dissipative rate. The condition $\lambda>2\gamma$ means the weak qubit-cavity coupling regime, while the condition $\lambda<2\gamma$ indicates the strong qubit-cavity coupling regime that the non-Markovian effect is very obviously\cite{Breuer2,H.M.Zou1,H.M.Zou2}.
We use the residue theorem to solve the Eq.(\ref{EB09}),  and the correlation function $F(t-t_{1})$ can be calculated as
\begin{eqnarray}\label{EB11}
F(t-t_{1})=\sum\limits_{i=1}^{4}F_{i}(t-t_{1}),
\end{eqnarray}
where
\begin{eqnarray}\label{EB12}
\begin{array}{l}

F_{1}(t-t_{1})=-\xi \exp\{[2\mu\beta t_{1}-2\tau_{0} \mu+(\mu\beta-\eta)(t-t_{1})]\}\\
F_{2}(t-t_{1})=\xi\exp\{[\mu\beta-\eta](t-t_{1})\}\\
F_{3}(t-t_{1})=\xi \exp\{-[\mu\beta+\eta](t-t_{1})\}\\
F_{4}(t-t_{1})=-\xi \exp\{-[2\mu\beta t_{1}-2\tau_{0}\mu+(\mu\beta+\eta)(t-t_{1})]\},

\end{array}
\end{eqnarray}
where $\xi=\frac{1}{8}\gamma\lambda$, $\mu=\lambda+i\omega_{0}$ and $\eta=\lambda+i(\omega_{D}+\omega_{L}-\omega_{k})$. In actual experiment, the evolution time $t$ is usually $t<1\times10^{2}s$ and $\beta=v/c\sim10^{-11}$\cite{Z.D.Liu,X-J.Liu}. According to the above conditions, we can known that $2\mu\beta t_{1}\rightarrow0$, under this condition, $\exp\{\pm2\mu\beta t_{1}\}=1$. So Eq.(\ref{EB12}) can be simplified as
\begin{eqnarray}\label{EB13}
\begin{array}{l}
\begin{aligned}
F(t-t_{1})=&\xi [(1-e^{-2\tau_{0}\mu})e^{(\mu\beta-\eta) (t-t_{1})}]\\
&+\xi[(1-e^{2\tau_{0}\mu})e^{-(\mu\beta+\eta) (t-t_{1})}].
\end{aligned}
\end{array}
\end{eqnarray}
Using the Laplace transform , Eq.(\ref{EB08}) becomes
\begin{eqnarray}\label{EB14}
\begin{array}{l}
\begin{aligned}
s\cdot C_{1}(s)+\xi(1-\frac{1}{b}) \frac{C_{1}(s)}{s+\varepsilon_{0}}+
\xi(1+b)\frac{C_{1}(s)}{s+\varepsilon_{1}}=1,
\end{aligned}
\end{array}
\end{eqnarray}
where $b=e^{2\tau_{0}\mu}$,  $\varepsilon_{0}=\eta-\mu\beta$ and $\varepsilon_{1}=\eta+\mu\beta$. 
Then the equation Eq.(\ref{EB14}) can be simplified as
\begin{eqnarray}\label{EB15}
\begin{array}{l}
\begin{aligned}
C_{1}(s)=\frac{b(s+\varepsilon_{0})(s+\varepsilon_{1})}{ bs^{3}+b(\varepsilon_{0}+\varepsilon_{1}) s^{2}+d_{1} s+d_{2}},
\end{aligned}
\end{array}
\end{eqnarray}
where $d_{1}=b\varepsilon_{0}\varepsilon_{1}-\xi(1-b)^{2}$ and  $d_{2}=\xi(b-1)(\varepsilon_{1}-b\varepsilon_{0})$.
We can get $C_{1}(t)$ by using the residue theorem,
\begin{eqnarray}\label{EB16}
\begin{array}{l}
C_{1}(t)=\sum\limits_{k=1}^{3}\frac{b(s_{k}+\varepsilon_{0})(s_{k}+\varepsilon_{1})}{3bs_{k}^{2}+2b(\varepsilon_{0}+\varepsilon_{1})s_{k}+d_{1}}e^{s_{k}t},
\end{array}
\end{eqnarray}
where $s_{k}(k=1,2,3)$ is the root of the equation $bs^{3}+b(\varepsilon_{0}+\varepsilon_{1}) s^{2}+d_{1} s+d_{2}=0$, the density matrix of the qubit in the basis$\{|E\rangle, |G\rangle\}$ at time $t$ reads
\begin{eqnarray}\label{EB17}
\rho(t)=\left(
\begin{array}{cc}
\rho_{EE}(0)|C_{1}(t)|^{2}&\rho_{EG}(0)C_{1}(t)\\
\rho_{GE}(0)C_{1}^{*}(t)&1-\rho_{EE}(0)|C_{1}(t)|^{2}
\end{array}
\right),
\end{eqnarray}
Taking the derivative of Eq.(\ref{EB17}), we get
\begin{eqnarray}\label{EB18}
\begin{array}{l}
\begin{aligned}
\frac{d}{dt}\rho(t)=&-i\frac{S(t)}{2}[\sigma_{+}\sigma_{-},\rho(t)]+\frac{\Gamma(t)}{2}[2\sigma_{-}\rho(t)\sigma_{+}\\
&-\sigma_{+}\sigma_{-}\rho(t)-\rho(t)\sigma_{+}\sigma_{-}].
\end{aligned}
\end{array}
\end{eqnarray}
This is the master equation for the reduced system dynamics. Obviously, $S(t)=-2\Im{[\dot{C}_{1}(t)/C_{1}(t)]}$ is a time dependent Lamb shift and $\Gamma(t)=-2\Re{[\dot{C}_{1}(t)/C_{1}(t)]}$ is a time dependent decay rate.
\section{Quantum speed limit and non-Markovianity}
\hspace{12pt}
In this section, we will briefly review the definitions of the QSLT and the non-Markovianity for an open quantum system. As a measure of statistical distance between quantum states, the Bures angle is defined as $B(\rho_{0},\rho_{t})=\arccos[F(\rho_{0},\rho_{t})]$, where $F(\rho_{0},\rho_{t})=\Tr[\sqrt{\sqrt{\rho_{0}}\rho_{t}\sqrt{\rho_{0}}}]$. In Ref.\cite{Deffner1}, the Bures angle was simplified as $B(\rho_{0},\rho_{t})=\arccos[\langle\psi_{0}|\rho_{t}|\psi_{0}\rangle]$ in open quantum systems. Here we will introduced the relative purity function to measure the trace distance\cite{F.Campaioli}, thus the Bures angle $B(\rho_{0},\rho_{t})$ can be written as
\begin{eqnarray}\label{EB19}
\begin{array}{l}
B(\rho_{0},\rho_{t})=\arccos(\sqrt{\frac{\Tr[\rho_{0}\rho_{t}]}{\Tr[\rho_{0}^{2}]}}).
\end{array}
\end{eqnarray}
Based on the von Neumann trace inequality and the Cauchy-Schwarz inequality, the QSLT is obtained as follows:
\begin{eqnarray}\label{EB20}
\begin{array}{l}
\tau_{qsl}=\max\{\frac{1}{\mathcal{V}_{\tau}^{op}},\frac{1}{\mathcal{V}_{\tau}^{tr}},\frac{1}{\mathcal{V}_{\tau}^{hs}}\}\sin^{2}[B(\rho_{0},\rho_{t})]\Tr[\rho_{0}^{2}],
\end{array}
\end{eqnarray}
where
$\mathcal{V}_{\tau}^{op,tr}=\frac{1}{\tau}\int_{0}^{\tau}dt||L_{t}(\rho_{t})||_{op,tr}$ and $\mathcal{V}_{\tau}^{hs}=\frac{1}{\tau}\int_{0}^{\tau}dt||L_{t}(\rho_{t})||_{hs}$. Owning to the relationship $\mathcal{V}_{\tau}^{op}\leq\mathcal{V}_{\tau}^{hs}\leq\mathcal{V}_{\tau}^{tr}$, the greater QSL velocity is $\mathcal{V}_{\tau}^{op}$ and QSLT bound is $\tau_{op}$. The QSLT in Eq.(\ref{EB20}) is the tightest bound,
\begin{eqnarray}\label{EB21}
\begin{array}{l}
\tau_{qsl} = \frac{1}{\mathcal{V}_{\tau}^{op}}\sin^{2}[B(\rho_{0},\rho_{t})]\Tr[\rho_{0}^{2}].
\end{array}
\end{eqnarray}
From Eq.(\ref{EB21}) and Eq.(\ref{EB17}), the QSLT is expressed as
\begin{eqnarray}\label{EB22}
\begin{array}{l}
\tau_{qsl} =\frac{1-|C_{1}(\tau)|^2}{(1/\tau)\int_{0}^{\tau}|\partial_{t}|C_{1}(t)|^2|dt},
\end{array}
\end{eqnarray}
is achieved when $|\psi(0)\rangle=|E\rangle$. where $C_{1}(t)$ has been calculated from Eq.(\ref{EB16}).

The non-Markovianity($\mathcal{N}$) is defined as the total backflow of information\cite{Y.J.Zhang,Baumgratz}
\begin{eqnarray}\label{EB23}
\begin{array}{l}
\mathcal{N} = \max\limits_{\rho_{1}(0),\rho_{2}(0)}\int_{\sigma>0}\sigma[t,\rho_{1}(0),\rho_{2}(0)]dt,
\end{array}
\end{eqnarray}
where $\sigma[t,\rho_{1}(0),\rho_{2}(0)]=\dot{D}[\rho_{1}(t),\rho_{2}(t)]$ is the change rate of the trace distance $D[\rho_{1}(t),\rho_{2}(t)]=\frac{1}{2} \Tr|\rho_{1}(t)-\rho_{2}(t)|$. When $\rho_{1}(0)=|E\rangle\langle E|$ and $\rho_{2}(0)=|G\rangle\langle G|$, the $\mathcal{N}$ in Eq.(\ref{EB23}) can obtain in the maximum value. $\mathcal{N}>0$ is non-Markovian process and $\mathcal{N}=0$ is Markovian process. The trace distance $D[\rho_{1}(t),\rho_{2}(t)]$ of the evolved states can be written as $D[\rho_{1}(t),\rho_{2}(t)]=|C_{1}(t)|^{2}$. Thus the $\mathcal{N}$ in Eq.(\ref{EB23}) can be rewritten as
\begin{eqnarray}\label{EB24}
\begin{array}{l}
\mathcal{N} = \frac{1}{2}[\int_{0}^{\tau}|\partial_{t} |C_{1}(t)|^{2}|dt+|C_{1}(\tau)|^{2}-1].
\end{array}
\end{eqnarray}
which connects to $\tau_{qsl}$ as 
\begin{eqnarray}\label{EB25}
\begin{array}{l}
\tau_{qsl}=\tau \frac{1-|C_{1}(\tau)|^{2}}{1-|C_{1}(\tau)|^{2}+2\mathcal{N}}.
\end{array}
\end{eqnarray}

\section{Results and discussion}
\begin{center}
	\includegraphics[width=4.2cm,height=3.5cm]{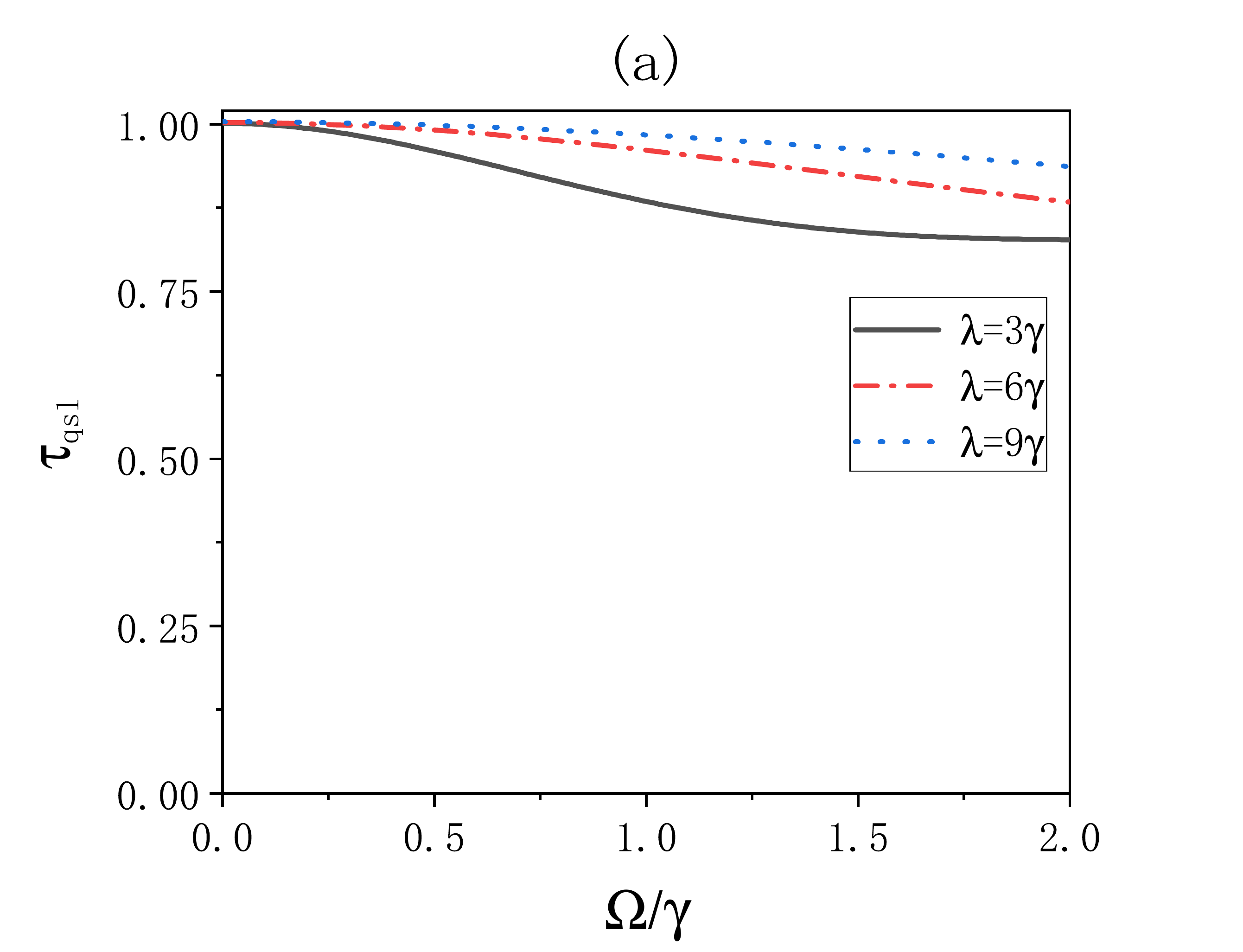}
	\includegraphics[width=4.2cm,height=3.5cm]{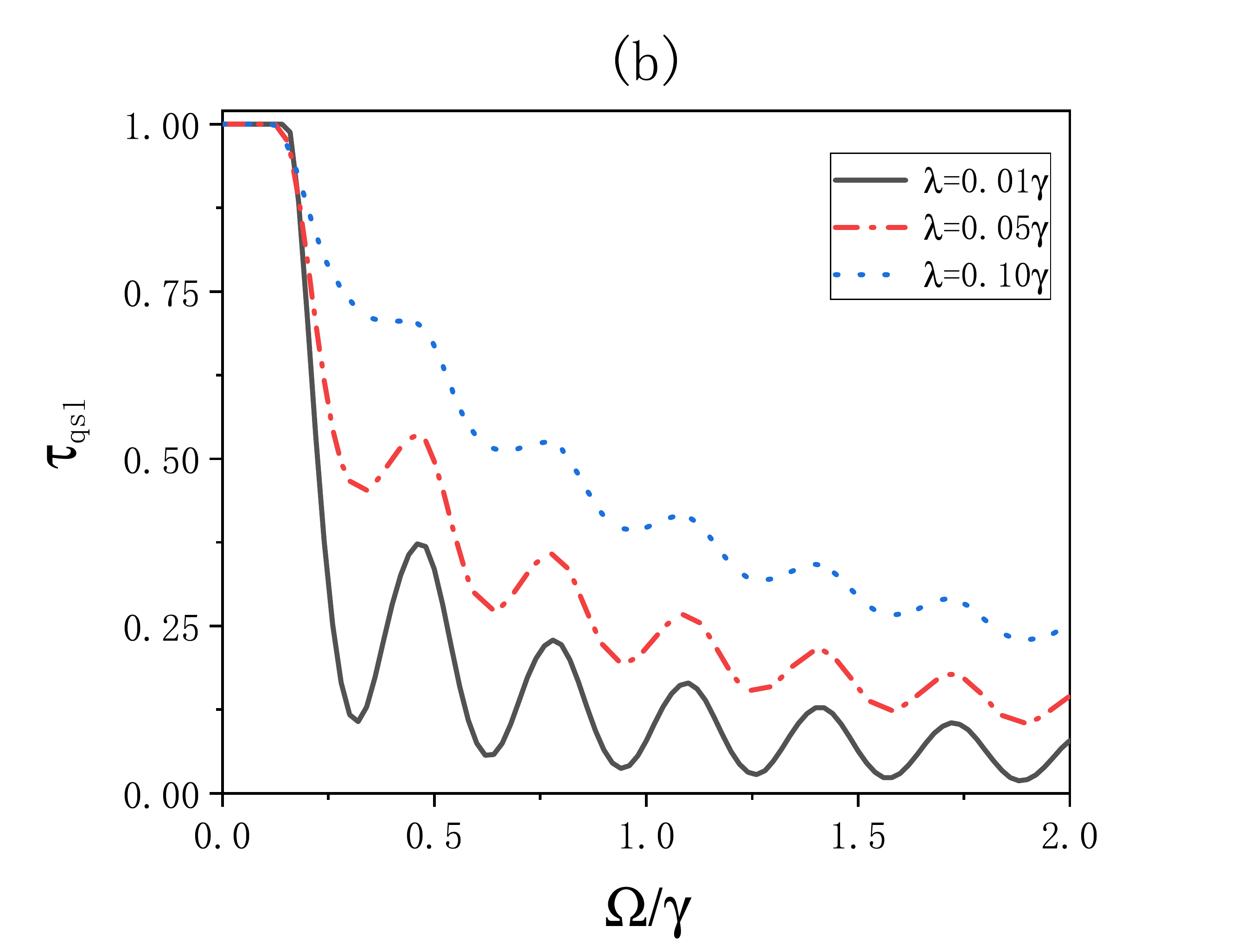}
	\parbox{8cm}{\small{\bf FIG.3.}
		$\tau_{qsl}$ as the function of driving strength $\Omega$.  (a) in the weak-coupling regime($\lambda>2\gamma$) and (b) in the strong-coupling regime($\lambda<2\gamma$). The velocity ratio $\beta=0$. The transition frequency $\omega_{0}=5.1\times10^{9}$. The dissipative rate $\gamma=10$. The actual evolution time $\tau=1$.}
\end{center}

Fig. 3 exhibits the variation curves of the $\tau_{qsl}$ with respect to the driving strength $\Omega$ when $\beta=0$ in the weak and strong coupling regimes, respectively. It is worth noting that Fig. 3(a) shows the significant speedup behavior can occur in quantum evolution process when the driving strength $\Omega$ reaches a certain critical value $\Omega_{c}$ in the weak-coupling regime. Namely, the system undergoes a standard evolution process when the driving strength $\Omega$ is less than the critical value $\Omega_{c}$, while the evolution can be accelerated very obviously when $\Omega>\Omega_{c}$. From Fig. 3(a), we see that, the smaller the value of $\lambda$ is, the smaller the critical value $\Omega_{c}$, the speedup phenomenon of the system evolution is more obvious at the same time. Fig. 3(b) shows the evolution curves of $\tau_{qsl}$ vs $\Omega$ in the strong-coupling regime. For different $\lambda$, the critical value $\Omega_{c}$ is same, but the smaller the value of $\lambda$ is, the more obvious the acceleration effect of quantum evolution. In addition, under the strong-coupling regime, the quantum evolution curve appears obviously collapse and recovery, this result well prove the information backflow under the non-Markovianity regime. Comparing Fig. 3(b) with Fig. 3(a), we find that the critical value $\Omega_{c}$ in the strong-coupling regime is significantly less than that in the weak-coupling regime. Thus both of the driving field and the strong-coupling can accelerate the quantum evolution.

\begin{center}
	\includegraphics[width=4.2cm,height=3.5cm]{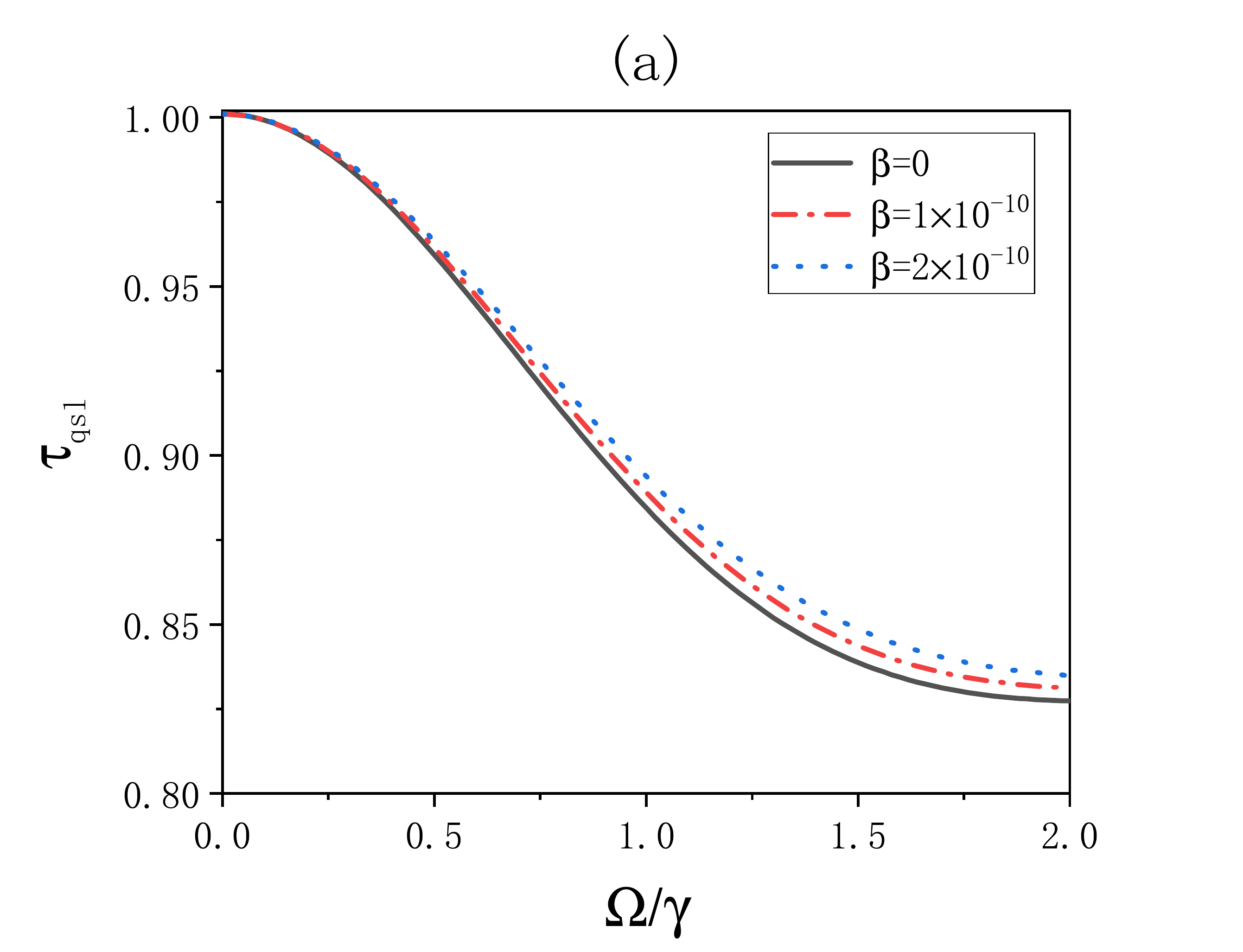}
	\includegraphics[width=4.2cm,height=3.5cm]{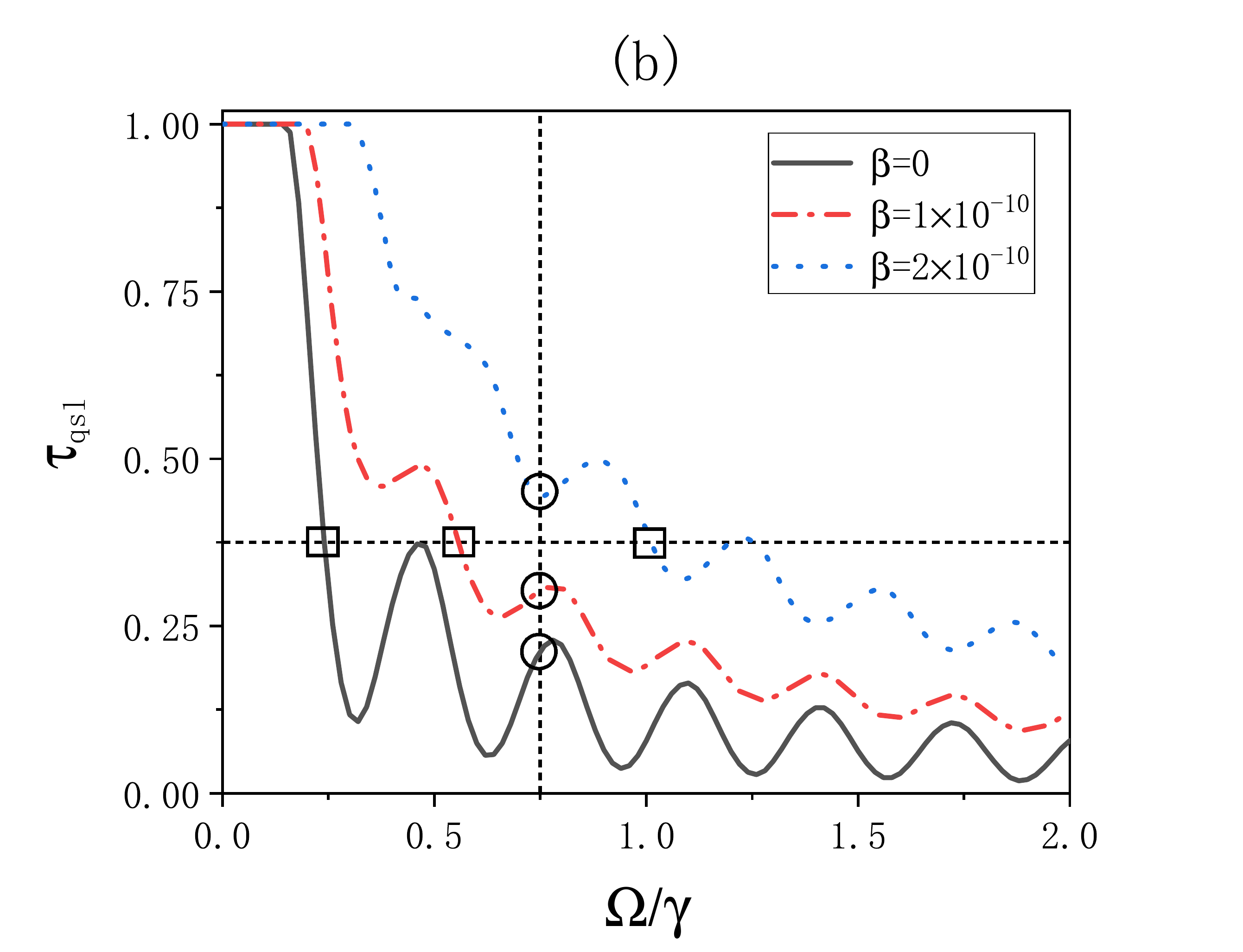}
	\parbox{8cm}{\small{\bf FIG.4.}
			$\tau_{qsl}$ as the function of driving strength $\Omega$.  (a) in the weak-coupling regime($\lambda=3\gamma$) and (b) in the strong-coupling regime($\lambda=0.01\gamma$). The transition frequency $\omega_{0}=5.1\times10^{9}$. The dissipative rate $\gamma=10$. The actual evolution time $\tau=1$.}
\end{center}

In order to know how the moving velocity and the classical field regulate the QSLT in the weak and the strong coupling regime, we give Fig. 4. Fig. 4(a) shows the QSLT as the function of the driving strength $\Omega$ in the weak-coupling regime with the actual evolution time $\tau=1$. It is worth noting that the transition form no speed-up to speed-up can occur  when the driving strength $\Omega$ reaches a certain critical value. Obviously, in the weak-coupling regime, $\Omega_{c}$ is tiny, while the quantum evolution can be accelerated when we increase driving strength $\Omega$.  We can also find that when we increase the qubit-velocity, the quantum evolution will become slower. 

However, in the strong-coupling regime, as shown in Fig. 4(b), the curve of $\tau_{qsl}$ exhibits very obvious oscillation. When $\Omega$ is very small, the qubit evolution at the actual time $\tau=1$. The evolution of the qubit will be accelerated when the $\Omega$ is bigger then a certain value. There are different critical values $\Omega_{c}$ when the qubit moves at different velocity. The larger $\beta$ value corresponds to a greater value of $\Omega_{c}$, and the larger $\beta$ value corresponds to the greater $\tau_{qsl}$ for a certain value of $\Omega$ as shown the circular mark in Fig. 4(b). Therefore, the qubit-velocity can delay the evolution of the quantum state. For the quantum with different velocity, we can obtain the same $\tau_{qsl}$ by selecting different value of $\Omega$ as shown the rectangle mark in Fig. 4(b).

\begin{center}
	\includegraphics[width=8.5cm,height=10cm]{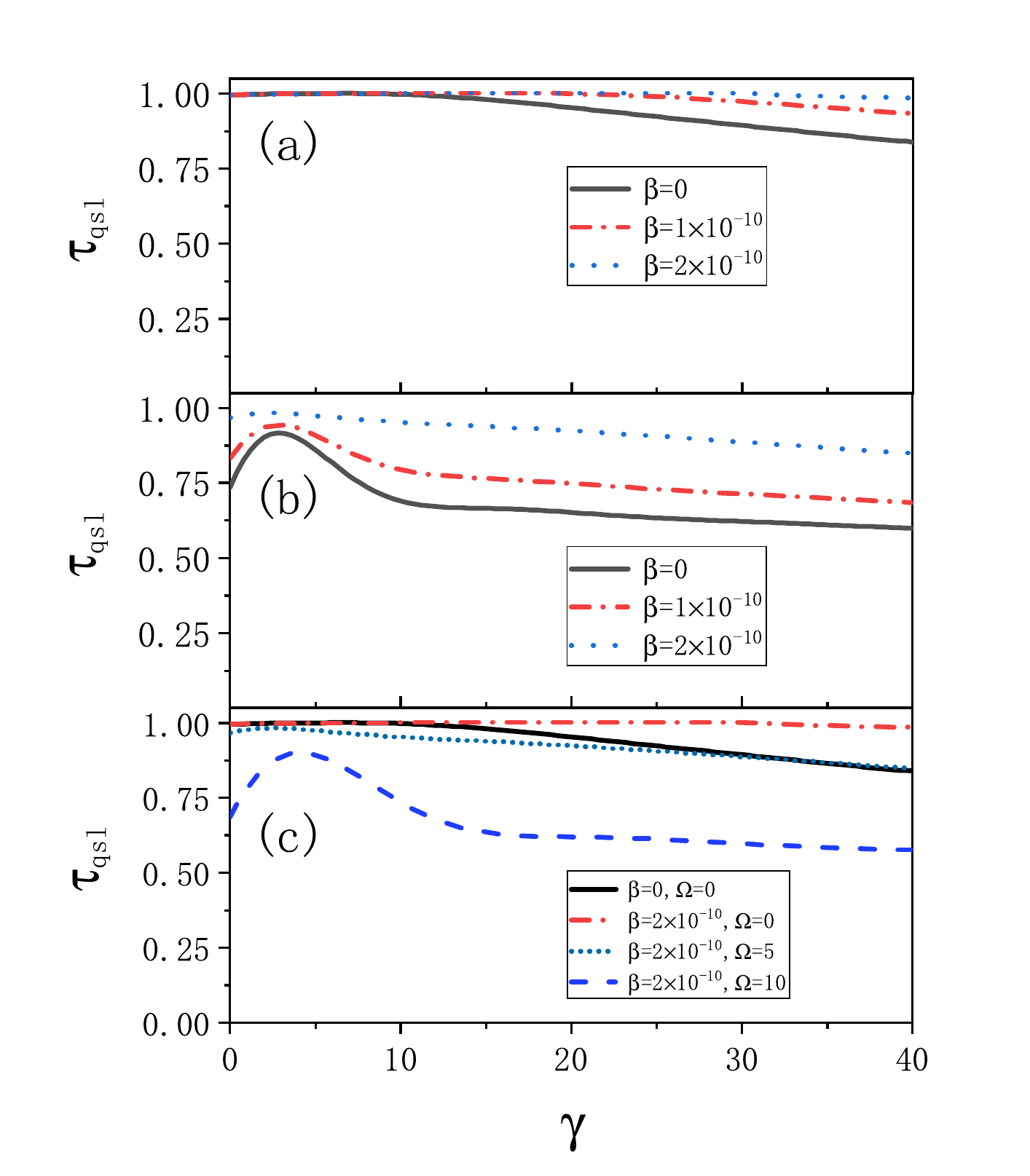}

	\parbox{8cm}{\small{\bf FIG.5.}
		$\tau_{qsl}$ as the function of dissipative rate $\gamma$. The driving strength $\Omega=0$ in (a), and $\Omega=5$ in (b). (c) is for different $\Omega$ and $\beta$.The transition frequency $\omega_{0}=5.1\times10^{9}$ . The spectral width parameter $\lambda=10$. The actual evolution time $\tau=1$.}
\end{center}

In Fig. 5, the curves of $\tau_{qsl}$ vs the dissipative rate are plotted for different velocity ratio when the driving strength $\Omega=0$ and $\Omega=5$, respectvely. Fig. 5(a) shows that without classical field, the qubit is always in the standard evolution if the dissipative rate $\gamma$ is less than the critical value $\gamma_{c}$, while the evolution can be speed-up when $\gamma >\gamma_{c}$. In addition, the critical value $\gamma_{c}$ gradually increases as the qubit-velocity become larger. This also indicates that the moving velocity of the qubit may play an important role in stabilizing quantum evolution. Fig. 5(b) exhibits the dependence of the $\tau_{qsl}$ on the dissipative rate $\gamma$ under the classical field. One can find that, in the presence of the classical field, there has been a significant acceleration evolution when the dissipation rate $\gamma$ is small. With the dissipative rate $\gamma$ increasing, the qubit will transit from a speedup evolution to recovery and collapse and then again undergo a speedup process. As the dissipative rate $\gamma$ gradually increases, the quantum evolution appears steadily acceleration phenomenon. When we increase the velocity of the qubit, we can find that QSLT will slowly down, this results is similar to Fig. 5(a).  

In Fig. 5(c) comparing the solid line and the dash dot line, we can find that the larger value of $\beta$ can delay the evolution of the qubit. Comparing the dash dot line and the dash line, we can know that the classical field can accelerate the revolution of the qubit. Comparing the solid line and the short dot line, we find that the curve of $\tau_{qsl}$ under $\beta=2\times 10^{-10}$ and $\Omega=5$ is close to that under $\beta=0$ and $\Omega=0$. Which indicates that the appropriate value of $\Omega$ can reduce the influence of the qubit-velocity on the QSLT. This result will provide a useful reference for the research of the cavity QED in theory and experiment.

\begin{center}
	\includegraphics[width=4.2cm,height=3.5cm]{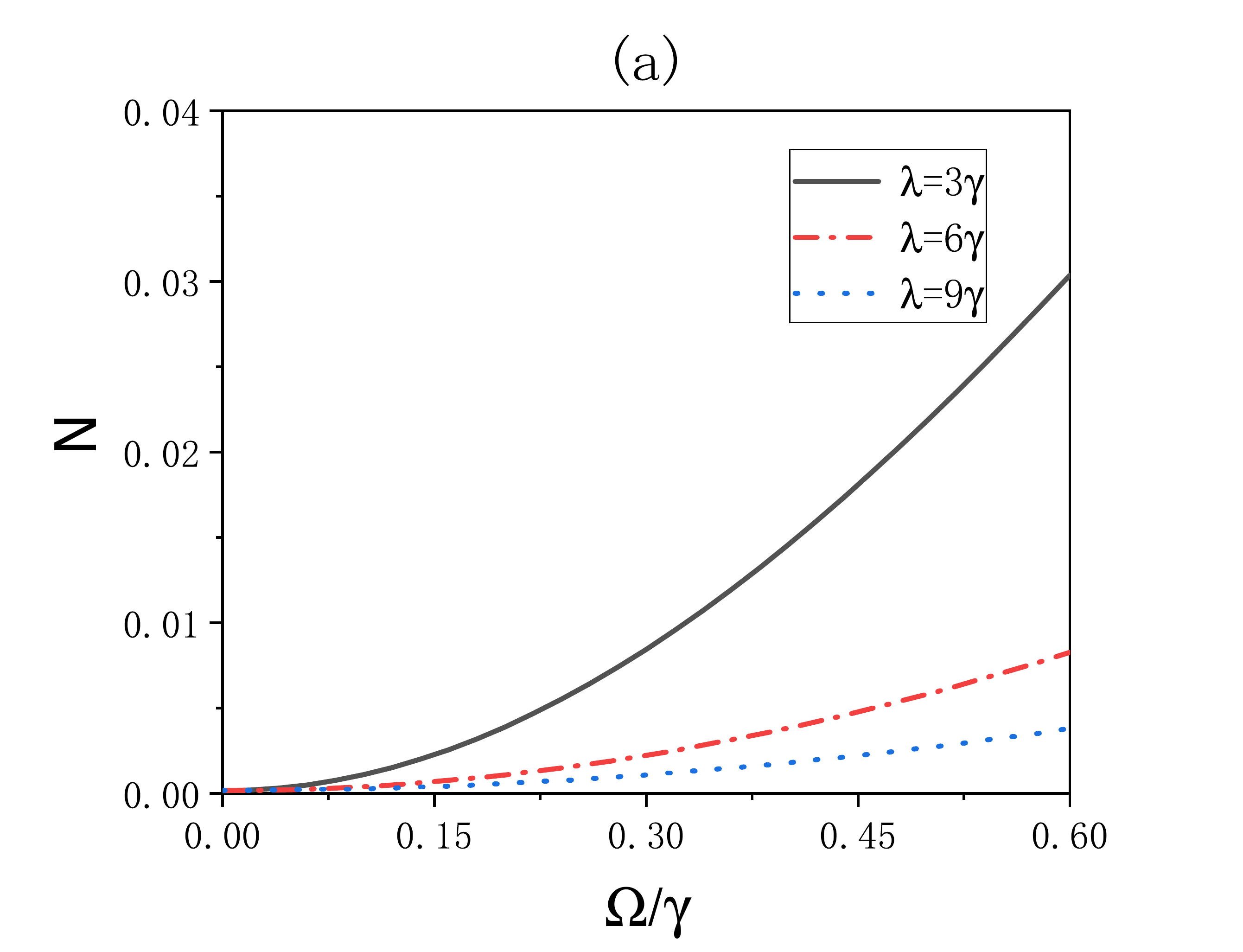}
	\includegraphics[width=4.2cm,height=3.5cm]{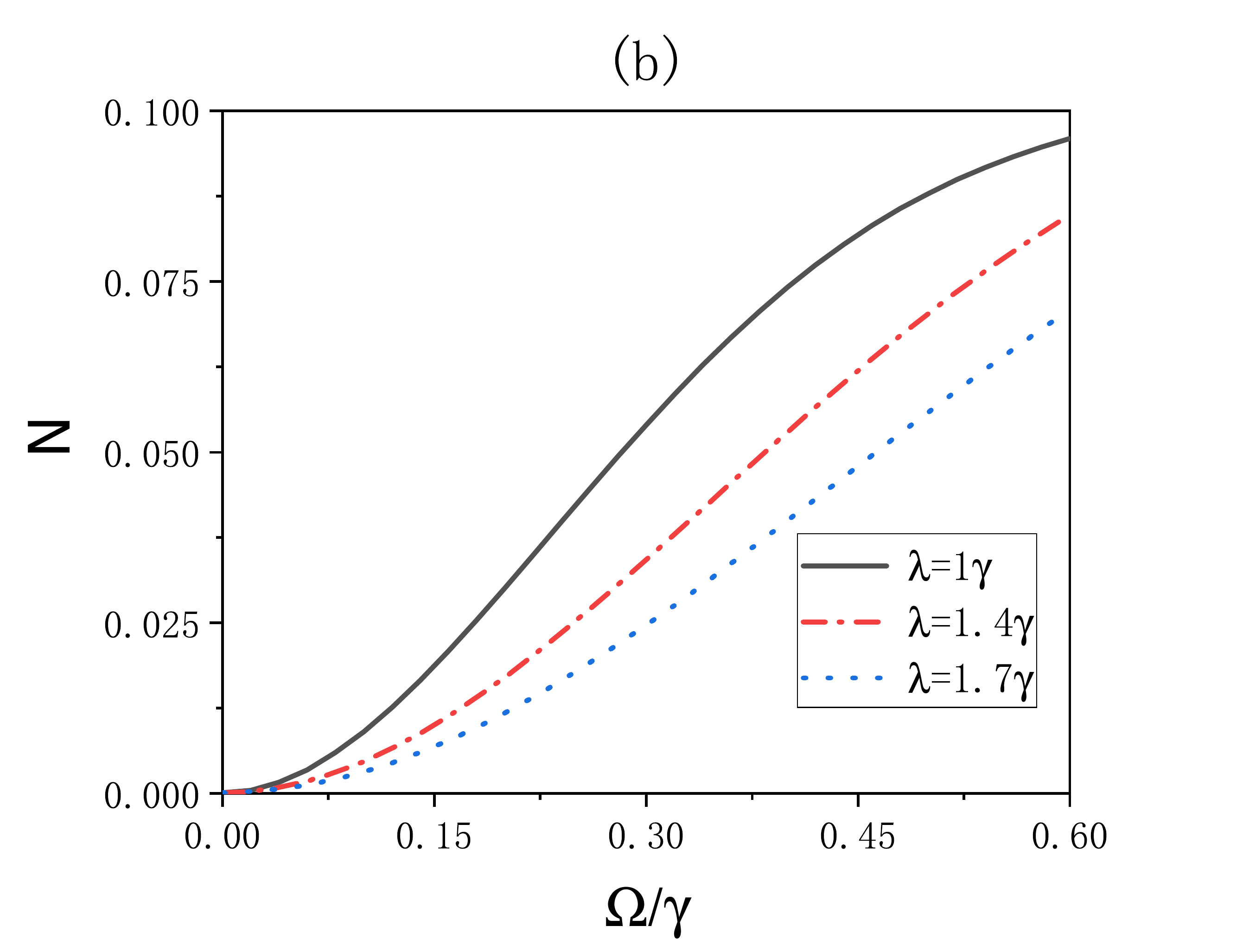}
	\parbox{8cm}{\small{\bf FIG.6.}
		$\mathcal{N}$ as a function of driving strength $\Omega$. (a) in the weak-coupling regime($\lambda>2\gamma$) and (b) in the strong-coupling regime($\lambda<2\gamma$). The velocity ratio $\beta=0$. The transition frequency $\omega_{0}=5.1\times10^{9}$. The dissipative rate $\gamma=10$. The actual evolution time $\tau=1$.}
\end{center}

\begin{center}
	\includegraphics[width=4.2cm,height=3.5cm]{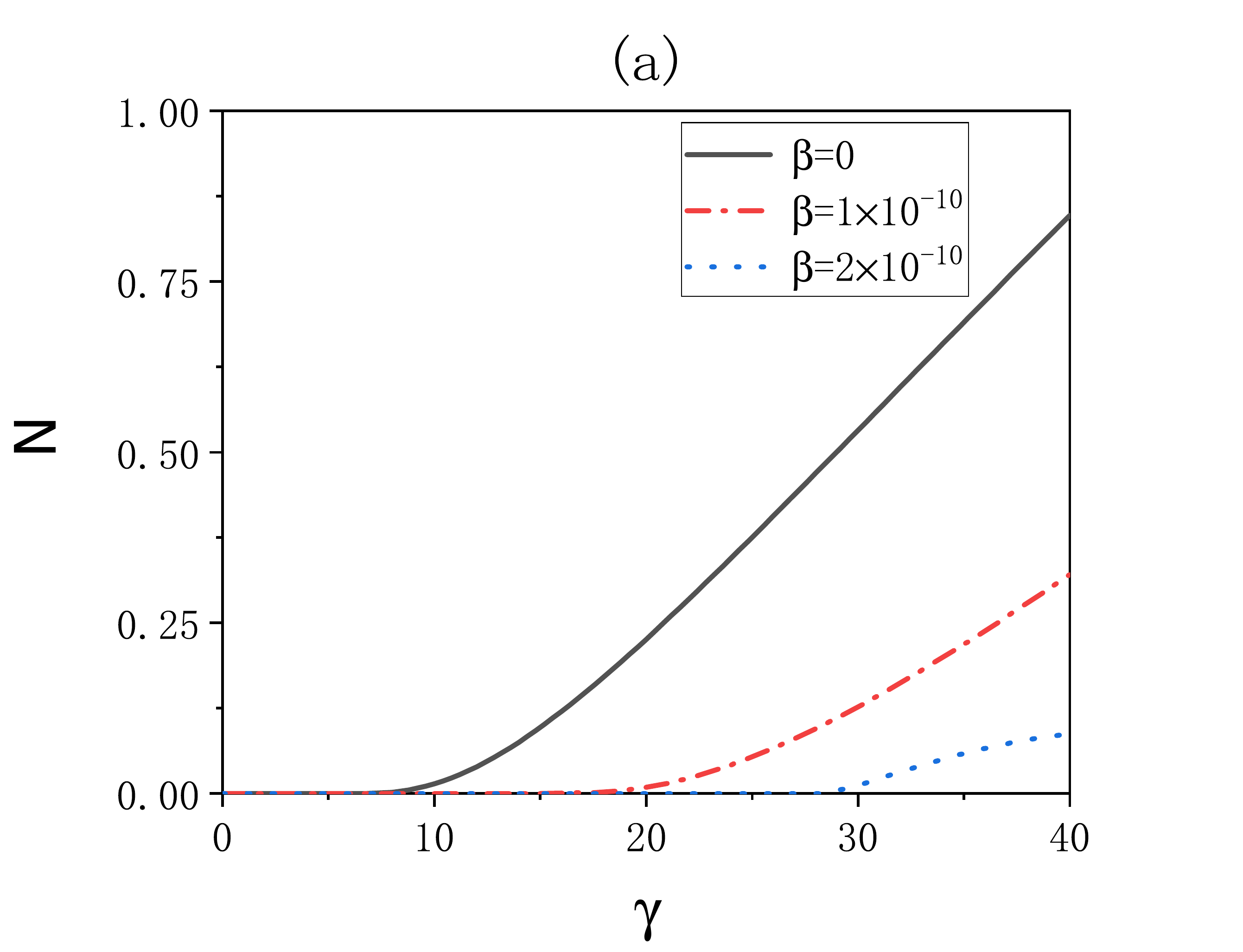}
	\includegraphics[width=4.2cm,height=3.5cm]{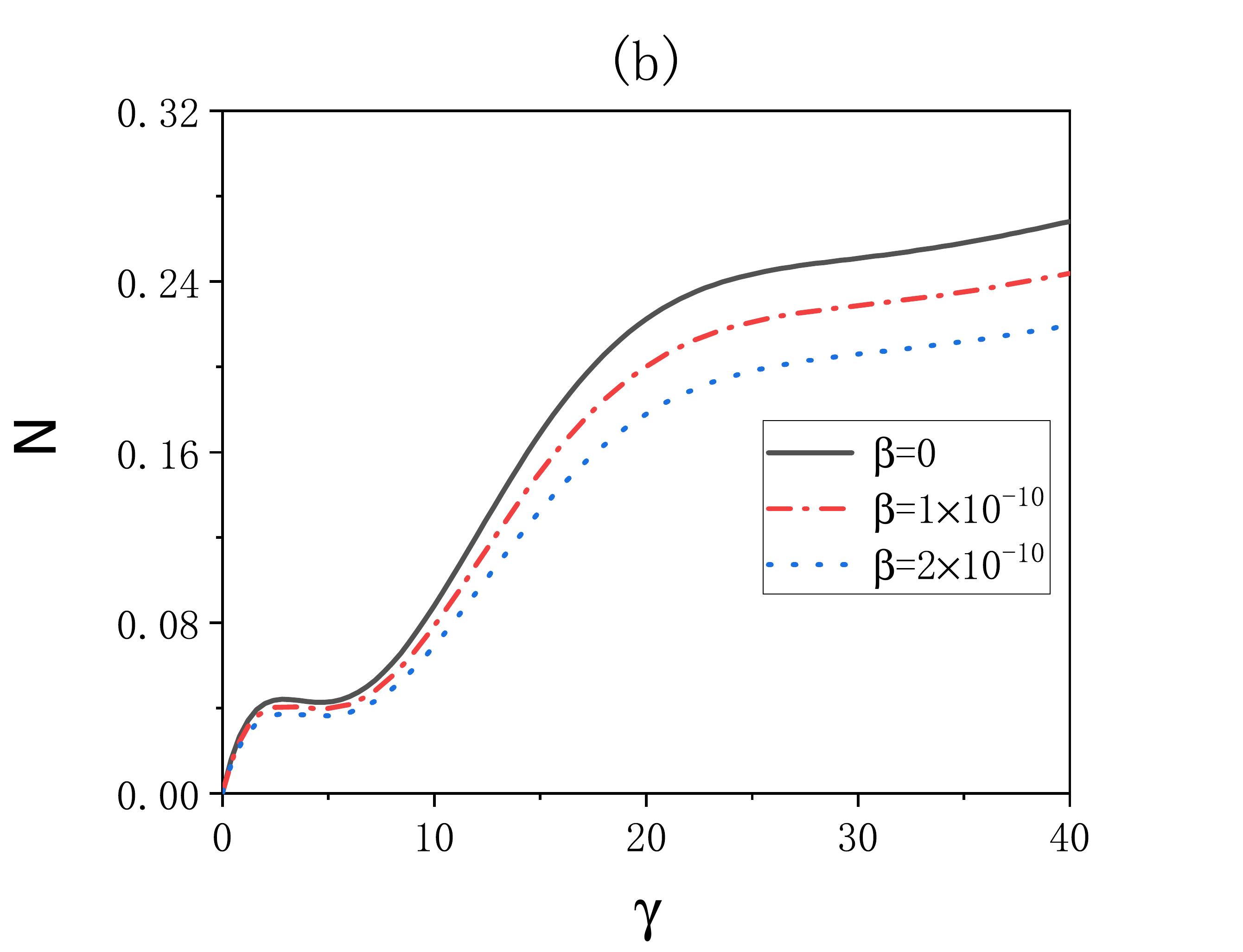}
	\parbox{8cm}{\small{\bf FIG.7.}
		$\mathcal{N}$ as the function of dissipative rate $\gamma$. The driving strength $\Omega=0$ in (a), and $\Omega=5$ in (b). The transition frequency $\omega_{0}=5.1\times10^{10}$. The spectral width parameter $\lambda=9$. The dissipative rate $\gamma=10$. The actual evolution time $\tau=1$.}
\end{center}

Fig. 6 exhibits the variation curves of non-Markovianity $\mathcal{N}$ with respect to driving strength $\Omega$ when $\beta=0$ in the weak and strong coupling regimes, respectively. Fig. 6(a) shows that, in the weak-coupling regime, the non-Markovianity $\mathcal{N}$ in the dynamics process is always equal to zero when $\Omega<\Omega_{c}$, and the non-Markovianity $\mathcal{N}$ increases monotonously with the driving strength $\Omega$ when $\Omega>\Omega_{c}$. When $\lambda$ takes different values, there is the diverse critical value $\Omega_{c}$ that $\mathcal{N}$ increases from zero. In addition, the smaller the value of $\lambda$ is, the more obvious non-Markovian characteristics. The dependence of $\mathcal{N}$ on $\Omega$ in the strong-coupling regime is shown in Fig. 6(b). We can see that, the smaller the value of $\lambda$ is, the smaller the critical value $\Omega_{c}$ is, the larger the $\mathcal{N}$ is. Besides, the critical value $\Omega_{c}$ in the strong-coupling regime is smaller than that in the weak-coupling regime. Thus both of the driving field and the strong-coupling can enhance the non-Markovianity $\mathcal{N}$ in the dynamics process.

In Fig. 7, we plot the non-Markovianity $\mathcal{N}$ against the dissipative rate $\gamma$ under the driving strength $\Omega=0$ in Fig. 7(a) and $\Omega=5$ in Fig. 7(b). Fig. 7(a) shows that the qubit undergoes the Markovian evolution process when the dissipative rate $\gamma$ is less than the critical dissipative rate $\gamma_{c}$. The qubit evolution will transit from the Markovian to the non-Markovian processes when $\gamma \geq\gamma_{c}$. In addition, the value of  $\gamma_{c}$ gradually becomes larger as $\beta$ increased. From Fig. 7(b), we can find that the non-Markovianity $\mathcal{N}$ will increase as $\gamma$ increases. $\beta$ has little effect on the $\mathcal{N}$ when $\gamma$ is very small, but the effect of $\beta$ on the $\mathcal{N}$ becomes very obvious and the larger value of $\beta$ corresponds to the smaller non-Markovianity $\mathcal{N}$.

\begin{center}
	\includegraphics[width=4.2cm,height=3.5cm]{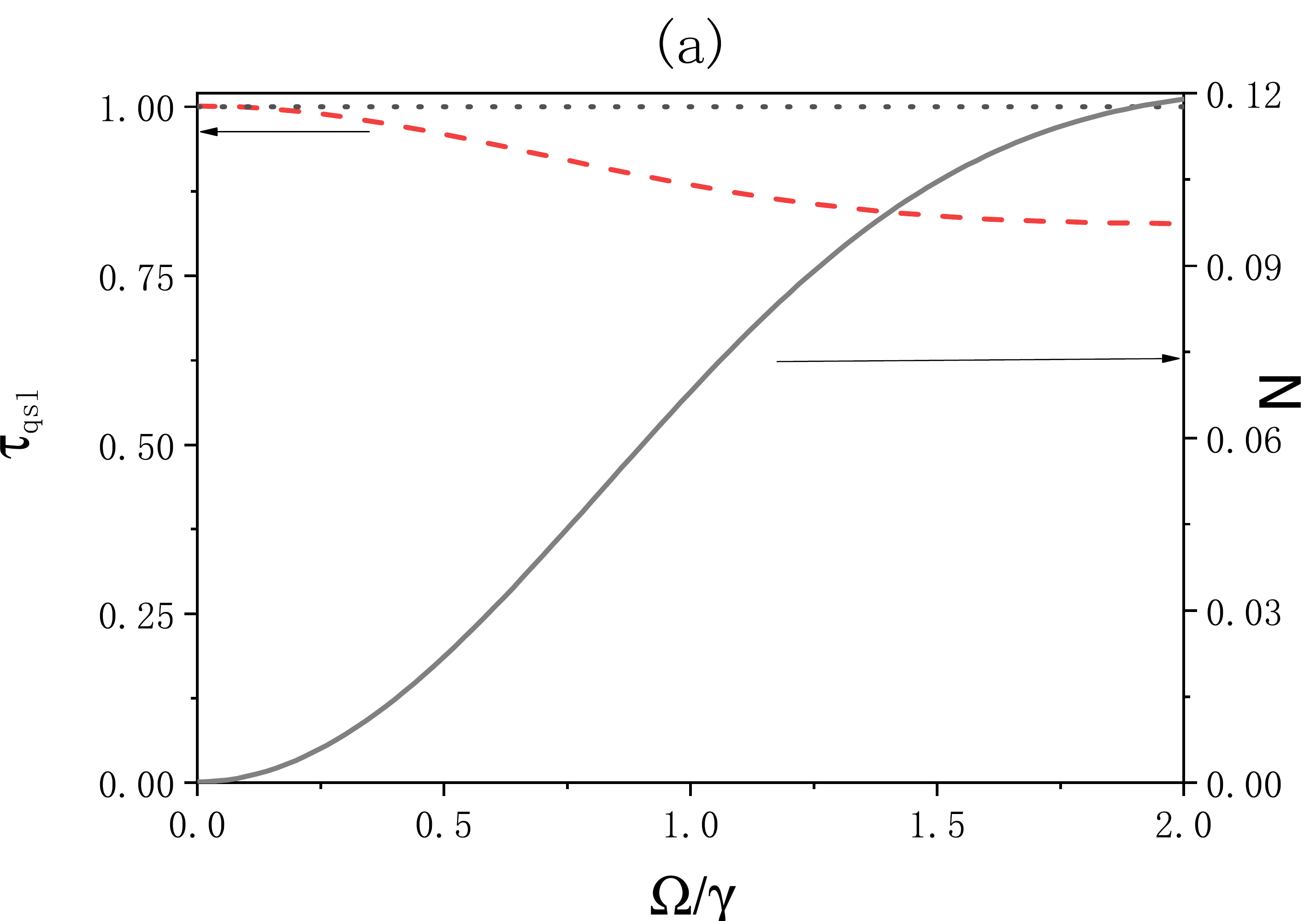}
	\includegraphics[width=4.2cm,height=3.5cm]{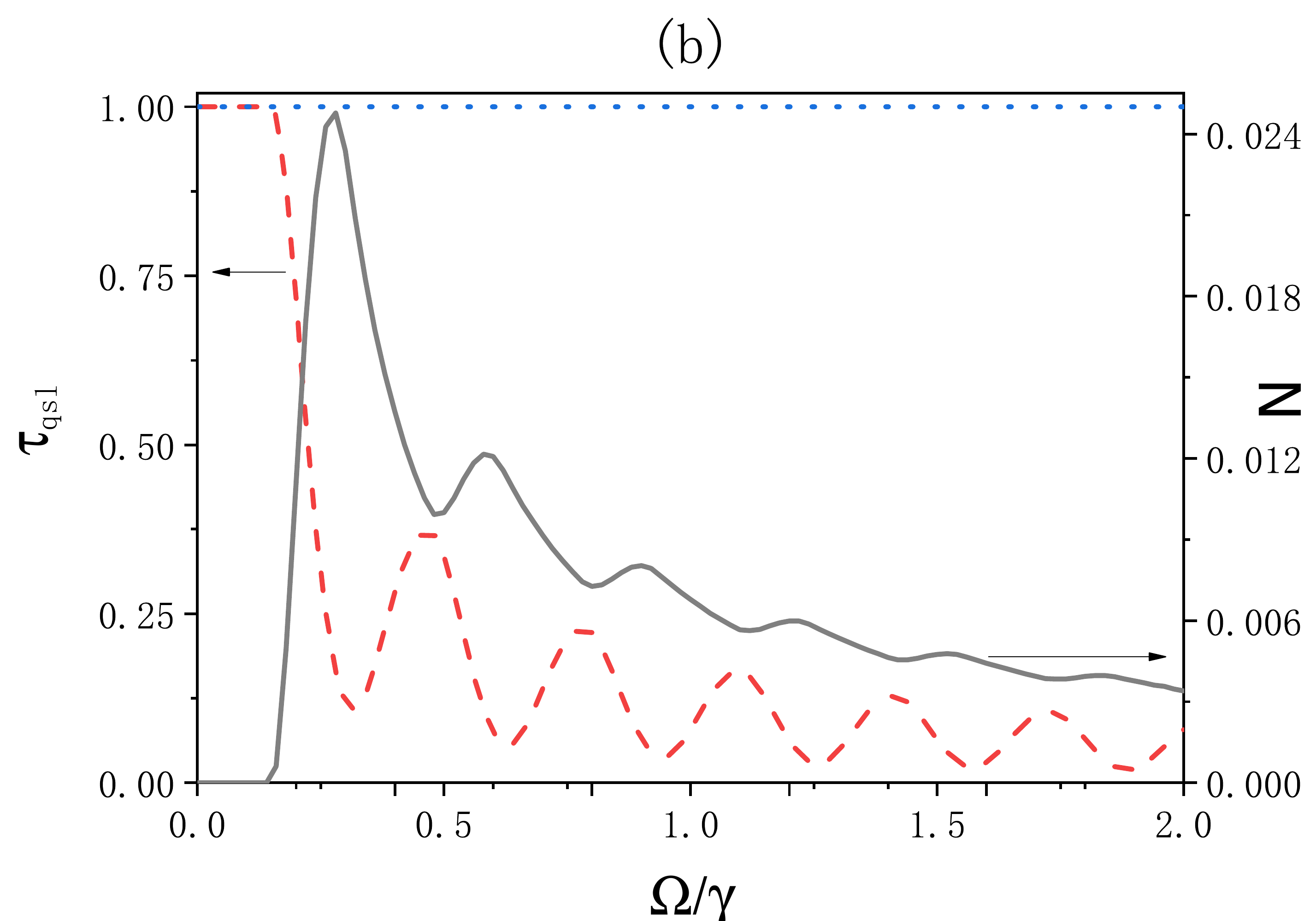}
	\parbox{8cm}{\small{\bf FIG.8.}
		$\tau_{qsl}$ and $\mathcal{N}$ as the functions of the driving strength $\Omega$. Here the spectral width parameter $\lambda=3\gamma$ in Fig. 8(a) and spectral width parameter $\lambda=0.01\gamma$ in Fig. 8(b).The dissipative rate $\gamma=10$. The velocity ratio $\beta=0$. The transition frequency $\omega_{0}=5.1\times10^{9}$. The dissipative rate $\gamma=10$. The actual evolution time $\tau=1$.}
\end{center}

\begin{center}
	\includegraphics[width=4.2cm,height=3.5cm]{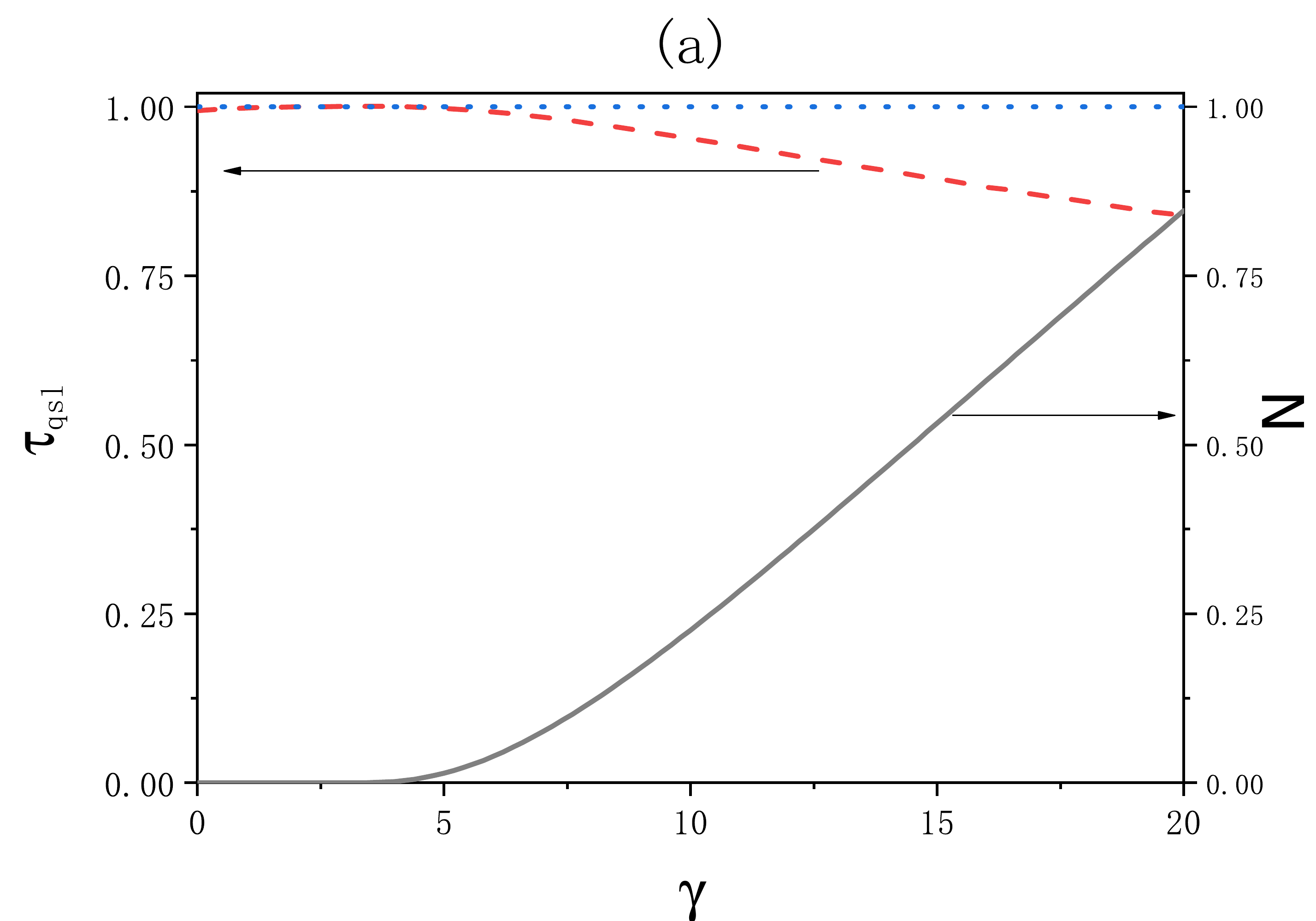}
	\includegraphics[width=4.2cm,height=3.5cm]{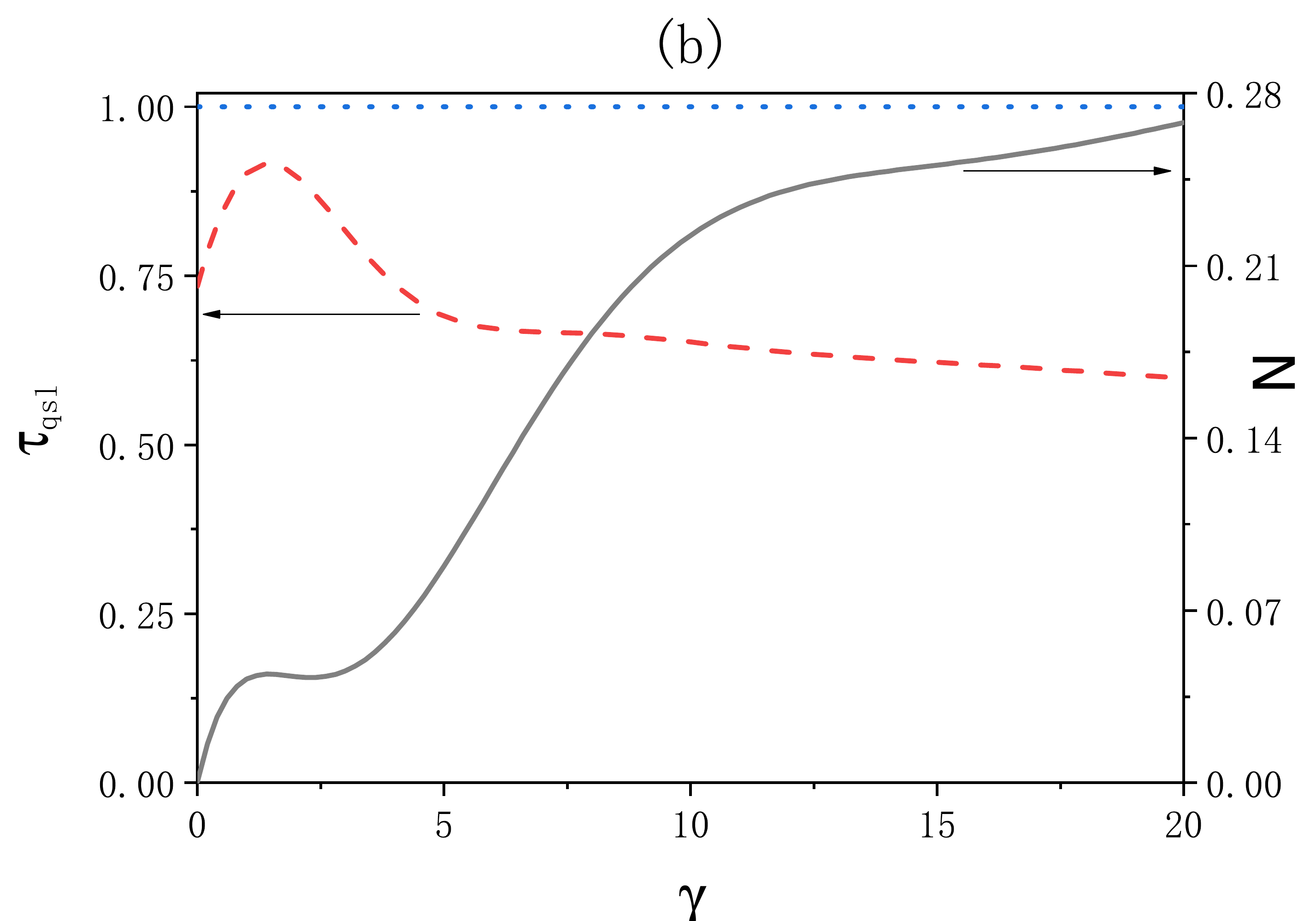}
	\parbox{8cm}{\small{\bf FIG.9.}
		$\tau_{qsl}$ and $\mathcal{N}$ as the functions of dissipative rate $\gamma$. Here the driving strength $\Omega=0$ in Fig. 9(a) and $\Omega=5$ in Fig. 9(b). The spectral width parameter $\lambda=10$. The velocity ratio $\beta=0$. The transition frequency $\omega_{0}=5.1\times10^{9}$. The actual evolution time $\tau=1$.}
\end{center}

In order to show more clearly the dependency relationship of the $\tau_{qsl}$ and the non-Markovianity, we draw Fig. 8 and Fig. 9.

Fig. 8 gives the curves of the $\tau_{qsl}$ and the non-Markovianity $\mathcal{N}$ with respect to the driving strength $\Omega$ when $\beta=0$ in the weak and strong coupling regimes, respectively. We can observe from Fig. 8(a) that, in the weak-coupling regime, $\mathcal{N}$ remains zero and $\tau_{qsl}$ stays at $1$ when $\Omega<\Omega_{c}$, but $\mathcal{N}$ will increase and $\tau_{qsl}$ experiences transition from no speed-up to speed-up evolution when $\Omega\geq\Omega_{c}$. Fig. 8(b) shows that, in the strong-coupling regime, the curves of $\tau_{qsl}$ and $\mathcal{N}$ appears obviously collapse and recovery. This trend can be correspond to Eq.(\ref{EB25}).

Fig. 9 exhibits the curves of the $\tau_{qsl}$ and non-Markovianity $\mathcal{N}$ against the dissipative rate $\gamma$ when the driving strength $\Omega=0$ and $\Omega=5$, respectvely. In Fig. 9(a), we can observe that, if $\Omega=0$, $\mathcal{N}=0$ and $\tau_{qsl}=1$ when $\gamma <\gamma_{c}$, while  $\mathcal{N}>0$ and $\tau_{qsl}<1$ when $\gamma >\gamma_{c}$. In Fig. 9(b), we can find that, if $\Omega=5$, there are $\mathcal{N}>0$ and $\tau_{qsl}<1$ when $\gamma$ is very small, $\tau_{qsl}$ and $\mathcal{N}$ will increase with $\gamma$ increasing. Then $\tau_{qsl}$ will always decrease and $\mathcal{N}$ will always enlarge as $\gamma$ increases. Namely, the larger value of  $\mathcal{N}$ corresponds to the smaller values of $\tau_{qsl}$.

From Fig. 8 and Fig. 9, the transition from Markovian to non-Markovian dynamics is the main physical reason of the quantum speed-up process, and both of the driving field and the strong-coupling can enhance the non-Markovianity in the dynamics process and speed up the evolution of the qubit.

\section{Conclusions}

In summary, we consider a model of a moving-qubit interacting with the multimode cavity, where the qubit is driven by the classical field. We obtain the analytic solution of the density operator of the qubit. We investigate the QSLT of the qubit and the non-Markovianity in the quantum process based on the classical field and the moving velocity of the qubit. The results show that the classical field also obviously accelerates the quantum evolution in both of the weak and the strong coupling regimes. Namely, the transition from Markovian to non-Markovian dynamics is induced by the classical field and the qubit-cavity coupling, and this transition is the main physical reason of the quantum speed-up process. Furthermore, the moving velocity of the qubit can decrease the non-Markovianity in the dynamics process and delay the evolution of the qubit,  and  the classical field can reduce the effect of the moving velocity of the qubit on the quantum evolution process.  This result shows that the classical field can  speed-up the quantum evolution of moving-qubit. Therefore, the controllable operation of quantum evolution can be realized by adjusting  the classical field strength, the qubit-reservoir coupling and the moving velocity of the qubit.  These results will provide a useful reference for the research of the cavity QED in theory and experiment.

\section*{Acknowledgments}
This work is supported by the National Natural Science Foundation of China (Grant No.11374096).

\end{document}